\documentstyle[11pt]{article}
\newcommand{\beq}{\begin{equation}}
\newcommand{\eeq}{\end{equation}}

\newcommand{\pdr}{\partial}
\newcommand{\beqs}{\begin{eqnarray}}
\newcommand{\eeqs}{\end{eqnarray}}

\newcommand{\nn}{\nonumber}

\begin{document}

\begin{center}
\vskip 2.5cm
{\LARGE \bf Propagators on the two-dimensional light-cone}

\vskip 1.0cm
{\Large  D.~K.~Park }
\\
{\large  Department of Physics  KyungNam University, Masan, 631-701
Korea}

\vskip 0.4cm
\end{center}

\centerline{\bf Abstract}

Light-cone quantization procedure recently presented is applied
to the two-dimensional light-cone theories. By introducing
the two distinct null planes it is shown that the modification
term in the two-dimensional massless light-cone propagators suggested
about twenty years ago vanishs.

\vfill

\newpage
\setcounter{footnote}{1}

\newcommand{\tr}{\;{\rm tr}\;}

In spite of the lack of manifest Lorentz covariance the light-cone
gauge (radiation gauge in light-cone coordinate) has been frequently
used for the calculation of perturbative QCD and the quantization
of the supersymmetric Yang-Mills theories.[1] Also it has been used
for the noncovariant formulation of string theories.[2]

However when the perturbation is calculated in the light-cone gauge,
there is a subtle point for the prescription of the "spurious"
singularity. If the usual principal-value(PV) prescription which
plays an important role in other non-covariant gauges is choosed,
the perturbation gives a poorly defined integral.[3] In order to
escape the difficulty the new prescription which is usually called
Mandelstam-Leibbrandt(ML) prescription[4] is suggested. Later the
ML-prescription is also derived in Ref.[5], in the framework of
equal-time canonical quantization and the renormalizibility of the
gauge theories formulated in this way is proved[6] in spite of the
appearance of an infinite number of nonlocal divergent term.

However this prescription is not directly applicable to calculation
based on the light-cone coordinate. Recently it is verified that the
ML-prescription is also derived in the light-cone coordinate when
one treats the relevant degrees of freedom carefully.[7]
In Ref[7] it was shown that in order to recover the ML-prescription
a characteristic surface $ x^+ = 0 $ must be used to initialize
the unphysical fields. In addition when the Poincar\'{e} charge is
calculated, not only the usual $ x^+ = 0 $ surface but also the
boundary wings must be included to the hypersurface where the
energy-momentum tensor is integrated.
More recently it is shown that two distinct
null planes are required for the implementation of the ML-prescription
in the interacting theories.[8]

About twenty years ago a similar problem with ML- and PV-prescriptions
was already discussed in the two-dimensional light-cone
coordinates.[9] The authors in Ref.[9]
suggested that the propagators must
be modified as follows in the two-dimensional light-cone by analyzing
the massless scalar and fermion theories
\beq
      \frac{1}{p^2 + i \epsilon} \Rightarrow
      \frac{1}{p^2 + i \epsilon} + \frac{i \pi}{2}
      \frac{\delta(p^+)}{\mid p^- \mid}.
\eeq
This suggestion is closely related with the choice of the
prescriptions in the light-cone gauge theories. To see this
by using
\beq
      ML\left( \frac{1}{p^+} \right) = PV \left( \frac{1}{p^+} \right)
                                  - i \pi \epsilon(p^-) \delta(p^+)
\eeq
where ML and PV mean the correspondent prescriptions, Eq.(1) is
simply written as
\beq
      \frac{1}{2 p^-} ML \left( \frac{1}{p^+} \right)
                  \Rightarrow
      \frac{1}{2 p^-} PV \left( \frac{1}{p^+} \right).
\eeq
This means that the "spurious" singularity problem occurs
not only in the gauge choice but also in the coordinate
choice. The purpose of this short paper is to apply the method
developed in Ref.[7, 8] to the two-dimensional massless theory
and to show that the modification term  in Eq.(1) is
cancelled by introducing the two distinct null planes. However
this does not guarantee that the ML-prescription is more
consistent than PV-prescription. The only thing we can say
is that these two prescriptions describe a different theories
in two dimension. Of course there can be various suggestion
to determine which is more consistent prescription in the
light-cone coordinates. The one way is to calculate the
vacuum expectation value(VEV) of a Wilson-loop operator
defined on the null-planes. Recently it is shown that
the result of VEV of Wilson-loop operator does not coincide
with that of the 't Hooft approach at two-loop level[10] if
the ML-prescription is choosed.[11] So it seems to be important
to perform the same calculation with the PV-prescription and
compare the result with that of the 't Hooft approach.

In order to show the disappearance of the modification term
appeared in Eq.(1), let us start with free scalar Lagrangian[12]
\beq
{\cal L} = \frac{1}{2} \pdr_{\mu} \phi \pdr^{\mu} \phi.
\eeq
Lagrangian (4) gives the equation of motion
\beq
\pdr_{+} \pdr_{-} \phi = 0
\eeq
and the energy-momentum tensor
\beqs
 T^{++} = \pdr_{-} \phi \pdr_{-} \phi, \\  \nn
 T^{--} = \pdr_{+} \phi \pdr_{+} \phi, \\  \nn
 T^{+-} = T^{-+} = 0.
\eeqs
If we take a Dirac procedure with a primary constraint
\beq
 \Omega = \Pi_{\phi} - \pdr_{-} \phi     \nn
\eeq
where $ \Pi_{\phi} $ is a canonical momentum of $\phi$ with
respect to the time coordinate $ x^+ $ and define a
Green's function as
\beq
 G(x, y) = i <0 \mid T^{+} \phi(x) \phi(y) \mid 0> \nn
\eeq
where $T^+$ is $x^+$ - ordered product, then we obtain
the result (1). In order to use the two distinct null
planes we consider the general solution of Eq.(5)
\beq
 \phi(x) = u(x^+) + v(x^-).
\eeq
If we add the boundary wings to the hypersurface $\sigma$
when the Poincar\'{e} charge is calculated by the formula
\beq
  P^{\mu} = \int_{\sigma} T^{\mu \nu} d\sigma_{\nu},
\eeq
the Poincar\'{e} charges become
\beqs
 P^+ = \int_{-\infty}^{\infty} dx^{-} \left[ \pdr_{-} v(x^-)
       \pdr_{-} v(x^-) \right]_{x^+ = 0}  \\  \nn
 P^- = \int_{-\infty}^{\infty} dx^{+} \left[ \pdr_{+} u(x^+)
       \pdr_{+} u(x^+) \right]_{x^- = 0}.
\eeqs
The difference of this approach from Ref.[9] is the non-zero
of $ P^-$. From the fact that the Heisenberg equation leads
the equations of motion
\beqs
        \pdr_{-} u = 0, \\  \nn
        \pdr_{+} v = 0,
\eeqs
we obtain the commutation relations
\beqs
         [u(x), u(y)]_{x^- = y^-} = - \frac{i}{4}
         \epsilon(x^+ - y^+) \\   \nn
         [v(x), v(y)]_{x^+ = y^+} = - \frac{i}{4}
         \epsilon(x^- - y^-)
\eeqs
where $\epsilon(x)$ is usual alternating function.
Equations of motion(12) and the commutation relation (13)
enable one to get the plane wave solution of $u(x)$ and
$v(x)$
\beqs
u(x^+) = \frac{1}{2 \sqrt{\pi}} \int_{-\infty}^{0} dp
         \frac{1}{\sqrt{\mid p \mid}}
         \left[ e^{i p x^+} a(p) + e^{-i p x^+} a^{\dagger}
         (p) \right]  \\ \\  \nn
v(x^-) = \frac{1}{2 \sqrt{\pi}} \int_{0}^{\infty} dp
         \frac{1}{\sqrt{p}} \left[e^{i p x^-} a^{\dagger}(p)
         + e^{-i p x^-} a(p) \right]
\eeqs
and the second quantization rules
\beqs
     [a(p), a^{\dagger}(q)] = \delta(p - q)  \\  \nn
     [a(p), a(q)] = [a^{\dagger}(p), a^{\dagger}(q)] = 0.
\eeqs
By inserting the plane wave solutions (14) to Eq.(11) the
Poincar\'{e} charges become
\beqs
    P^+ = \frac{1}{2} \int_{0}^{\infty} dp p [a^{\dagger}(p)
    a(p) + a(p) a^{\dagger}(p)]  \\  \nn
    P^- = \frac{1}{2} \int_{-\infty}^{0} dp \mid p \mid
          [a^{\dagger}(p) a(p) + a(p) a^{\dagger}(p)].
\eeqs
 From Eq.(16) we can obtain the Hamiltonian and momentum
operators derived in usual Minkowski space. This is the
evidence of the isomorphism between the light-cone
quantized theory and the equal-time quantized theory
commented in Ref.[13].
In order to obtain the propagators we define the Green's
function
\beq
g(x - y) = i <0 \mid T^{\ast} \phi(x) \phi(y) \mid 0>.
\eeq
where $ T^{\ast} $-ordering is taken with respect to the
$ x^+ $ coordinate for $v$ and $ x^- $ coordinate for
$ u $. If one uses the integral representation of the
step function
\beq
\theta(x) = \frac{1}{2 \pi i} \int_{-\infty}^{\infty}
            d\tau \frac{e^{i x \tau}}{\tau - i \epsilon},
\eeq
the direct calculation gives
\beq
g(x) = - \frac{2}{(2 \pi)^2}
       \int_{-\infty}^{\infty}dp^+
       \int_{-\infty}^{\infty}dp^-
       \frac{1}{2 p^+ p^- + i \epsilon}
       e^{i(p^+ x^- + p^- x^+)}.
\eeq
So the modification term appeared in Eq.(1) disappears from the
propagator. This result seems natural since the light-cone
quantization procedure formulated in this way
gives a Causal-prescription in four-dimensional
light-cone gauge theories[7] and in the two-dimensional
massless case the Causal-prescription is manifestly usual ML-
prescription.

For the completeness let us consider the fermionic case too.
For fermionic case the equations of motion are
\beqs
\pdr_{+} \psi_+ = 0  \\  \nn
\pdr_{-} \psi_- = 0
\eeqs
where $\gamma$ matrices are taken as
\beqs
\gamma^0 = \left( \begin{array}{cc}
                  0  &  1  \\
                  1  &  0
                  \end{array} \right) \hspace{.1in}
\gamma^1 = \left( \begin{array}{cc}
                  0  &  1  \\
                  -1 &  0
                  \end{array} \right) \hspace{.1in}
\gamma^5 = \gamma^0 \gamma^1 = \left( \begin{array}{cc}
                                      -1 &  0  \\
                                      0  &  1
                                      \end{array} \right) \\ \nn
\gamma^+ = \sqrt{2} \left( \begin{array}{cc}
                           0  &  1  \\
                           0  &  0
                           \end{array} \right) \hspace{.1in}
\gamma^- = \sqrt{2} \left( \begin{array}{cc}
                           0  &  0  \\
                           1  &  0
                           \end{array} \right).
\eeqs
Since the null-plane formalism gives a quantization rule
\beqs
 \{ \psi^{\ast}_{+}(x), \psi_{+}(y) \}_{x^+ = y^+} = \delta(x^- - y^-)  \\
 \{ \psi^{\ast}_{-}(x), \psi_{-}(y) \}_{x^- = y^-} = \delta(x^+ - y^+),
\eeqs
one can derive the plane-wave solutions
\beqs
 \psi_{+}(x^-) = \frac{1}{\sqrt{2 \pi}}
                 \int_{0}^{\infty}dk \left[
                 a(k) e^{-ikx^-} + b^{\dagger}(k) e^{ikx^-}
                                     \right]   \\
 \psi_{-}(x^+) = \frac{1}{\sqrt{2 \pi}}
                 \int_{-\infty}^{0} dk \left[
                 a(k) e^{-ikx^+} + b^{\dagger}(k) e^{ikx^+}
                                     \right]
\eeqs
,which is already used in Ref.[13] when proving the isomorphism
between the light-cone and equal-time quantized theories, and the
second quantization rules
\beq
\{a(p), a^{\dagger}(q) \} = \{b(p), b^{\dagger}(q) \}
 = \delta(p - q).
\eeq
By defining the Green's function
\beq
 g_{\alpha \beta}(x - y) = i <0 \mid
 T^* \psi_{\alpha}(x) \bar{\psi_{\beta}}(y) \mid 0>,
\eeq
the direct calculation gives
\beq
g(x - y) = \frac{\sqrt{2}}{(2 \pi)^2}
           \int_{-\infty}^{\infty}dk^+dk^-
           \frac{\gamma^{\mu} k_{\mu}}{k^2 + i \epsilon}
           e^{i k (x - y)}
\eeq
which is the usual Causal-prescription Green's function.
So it is shown that the modification term  in
two-dimensional massless propagators can be cancelled by
applying the light-cone quantization
procedure recently developed and introducing the
two distinct null planes. The appearance and the disappearanec
of the modification term is closely related with a recent
debate between ML- and PV-prescriptions which occured in
four-dimensional light-cone gauge theory. As was stated
previously, the disappearance of the modification term
by introducing the two distinct null planes does not
guarantee that ML-prescription is more consistent than
PV-prescription. To determine which is more consistent
in two-dimensional theories it is important to calculate
the VEV of the Wilson-loop operator defined at the
null planes as suggested before. This will be
reported elsewhere.

\bigskip

\noindent {\em ACKNOWLEDGMENTS}

This work was carried out with support from the Korean Science and
Engineering Foundation.

\end{document}